\documentclass[twocolumn,showpacs,aps]{revtex4}
\usepackage{amssymb}
\usepackage{epsfig}
\usepackage{amsfonts}
\usepackage{amsmath}
\usepackage{amssymb}
\usepackage{graphicx}

\begin{document}

\title[Self-Guiding of Electromagnetic Beams in Degenerate E-P Plasma]{Self-Guiding of
Electromagnetic Beams in Degenerate Relativistic Electron-Positron
Plasma}
\author{V.I. Berezhiani$^{1,2}$, N.L. Shatashvili$^{1,3}$}
\affiliation{$^{1}$Andronikashvili Institute of Physics, TSU,
Tbilisi \ 0177, Georgia}
\affiliation{$^{2}$School of Physics,
Free University of Tbilisi, Georgia}
\affiliation{$^{3}$Department
of Physics, Faculty of Exact and Natural Sciences, Ivane
Javakhishvili Tbilisi State University, TSU, Tbilisi 0179,
Georgia}

\pacs{42.65.Jx, 52.27.Ny, 52.35.Mw, 52.38.Hb, 47.75.+f}

\begin{abstract}
The possibility of self-trapped propagation of electromagnetic
beams in the fully degenerate relativistic electron-positron
plasma has been studied applying Fluid-Maxwell model; it is shown
that dynamics of such beams can be described by the generalized
Nonlinear Schr\"odinger equation with specific type of saturating
nonlinearity. Existence of radially symmetric localized solitary
structures is demonstrated. It is found that stable solitary
structures exist for the arbitrary level of degeneracy.

\end{abstract}

\pacs{42.65.Jx, 52.27.Ny, 52.35.Mw, 52.38.Hb, 47.75.+f}

\maketitle


Recent observations as well as the modern theoretical
considerations indicate on the existence of super-dense
electron-positron plasmas in variety of astrophysical
environments. The presence of e-p plasma is also argued in the MeV
epoch of the early Universe \cite{bib:Tajima}. Intense e-p pair
creation takes place during the process of gravitational collapse
of massive stars \cite{bib:Stenflo}; it is shown that in certain
circumstances the gravitational collapse of the stars may lead to
the charge separation with the field strength exceeding the
Schwinger limit resulting in e-p pair plasma creation with
estimated density to be $\sim 10^{34}\,cm^{-3}$ \
\cite{bib:ruffini}. The e-p plasma density can be in the range
$(10^{30}\div 10^{37})\,cm^{-3}$ of the GRB source
\cite{bib:aksenov}.

For highly compressed state the plasma behaves as a degenerate
Fermi gas provided that averaged inter-particle distance is
smaller than the thermal de Broglie wavelength. As the density
increases Fermi energy of the particles \
$\epsilon_F^{\pm}={\hbar}^2\,(3{\pi}^2\,n^{\pm})^{2/3}/{2m_e}$ \
with \ $+ (-)$ \ accounting for positrons and electrons,
respectively, becomes larger than the interaction energy \ ($\sim
e^2(n_0^{\pm})^{1/3}$) \ and a mutual interaction of the plasma
particles becomes unimportant -- plasma becomes more ideal
\cite{bib:Landau}. This condition is fulfilled for a sufficiently
dense fluid when \ $n_0^{\pm}\gg
(2m_e\,e^2/(3\pi^2)^{2/3}{\hbar}^2)^3=6.3\cdot 10^{22}\,cm^{-3}$ .
When density increases further particle's relativistic motion
shall be taken into account leading to the relativistic Fermi
energy of the particles in the following form \ $\epsilon^{\pm}
_{F} = m_{e}c^{2}\,\left[ \left( 1+(R^{\pm})^{2}\right)
^{1/2}-1\right]$ , where \ $R^{\pm}=p^{\pm}_{F}/m_{e}c$, \
$p^{\pm}_{F}$ \ -- is the Fermi momentum which is related to the
rest-frame particle density by the following relation \
$p^{\pm}_{F} = m_{e}c\,\left( n^{\pm}/n_{c}\right)^{1/3}$, here \
$n_{c} = 5.9\times 10^{29}\,cm^{-3}$ \ is the normalizing critical
number-density \cite{bib:BST_deg}. Thus, when \ $n^{\pm}\gg n_c$ \
plasma turns out to be ultra-relativistic even for
non-relativistic temperature \ $T^{\pm}\ll \epsilon_F^{\pm}$ .
Here we would like to emphasize that the pair plasmas with such
densities can not be in complete thermodynamic equilibrium with
the photon gas into which it annihilates \cite{bib:Katz}.
Equilibrium is reached within the time-period related mainly to
the electron-positron annihilations. Subsequently, the
thermodynamic equilibrium between pairs and photons (with zero
chemical potential) will be achieved. Plasma becomes optically
thick with steady state pair density defined by plasma
temperature. On the other hand for high density optically thin
plasma, as it was shown in \cite{bib:BST_deg}, the e-p
annihilation time becomes considerably short \ ($\tau_{ann}\approx
0.3\times 10^{-16}\,sec$ \ for the densities \ $\sim
10^{30}\,cm^{-3}$). However, it is still larger than the
corresponding densities plasma oscillations characteristic
time-scale \ $[\sim \omega_{e}^{-1}]$; for instance, for the
densities \ $(10^{30} \div 10^{35})\,cm^{-3}$ \ we find that \
$(\tau_{ann}\omega_{e}^{-1})$ \ is in the range \ $(60 \div 7)$
\cite{bib:BST_deg} and collective plasma high frequency
oscillations have enough time to manifest themselves.

Generation of high density e-p plasma is presumably augmented by
production of intense pulses of X-- and Gamma--rays. To understand
the dynamics of such pulses emanating from the compact
astrophysical objects as well as to study the nonlinear
interactions of intense laser pulses and dense degenerate plasmas
it is important to investigate the wave self-modulation and
soliton formation phenomena in dense e-p plasmas. The existence of
stable localized envelope solitons of electromagnetic (EM)
radiation has been suggested as a potential mechanism for the
production of micro-pulses in AGN and pulsars \cite{bib:kennel}.
Localized solitons created in the plasma dominated era are also
invoked to explain the observed inhomogeneities of the visible
universe \cite{bib:Tajima}. Recently, the existence of
soliton-like electromagnetic distributions in a fully degenerate
electron-positron plasma was shown in \cite{bib:BST_deg} applying
relativistic hydrodynamic and Maxwell equations. For circularly
polarized wave it was found that the soliton solutions exist both
in relativistic as well as nonrelativistic degenerate plasmas and
the possibility of plasma cavitation was also shown.

\bigskip

In the present paper we apply the Fluid-Maxwell model presented in
\cite{bib:Stenflo,bib:Haas,bib:BST_deg} to investigate the
possibility of self-trapping of intense electromagnetic pulse in
transparent degenerate electron-positron (e-p) plasma in a limit
of narrow pulse \ $L_{\perp}\ll L_{\|}$ \ (where $L_{\|}$ and
$L_{\perp}$ are the characteristic longitudinal and transverse
spatial dimensions of the field, respectively) to demonstrate the
formation of stable 2D solitonic structures in such plasma.

Basic set of Maxwell-Fluid equations describing the dynamics of
unmagnetized fully degenerate electron-positron plasma reads as
\cite{bib:BST_deg}:
\begin{equation}
\frac{\partial^{2}{\bf A}}{\partial t^{2}}-c^{2}\Delta{\bf A} +
c\, \frac{\partial}{\partial t}\left( {\bf \nabla}\varphi\right) -
4\pi e c\,(N^{+}{\bf V}^{+} - N^{-}{\bf V}^{-})=0 \ , \label{B8}
\end{equation}
\begin{equation}
\Delta\varphi = - 4\pi e\,(N^{+} - N^{-}) \ , \label{B9}
\end{equation}
\begin{equation}
\frac{\partial}{\partial t}\left( G^{\pm}{\bf p}^{\pm} \pm
\frac{e}{c}{\bf A}\right) +{\bf \nabla}\left(
m_{e}c^{2}\,G^{\pm}\,\gamma \pm e\,\varphi\right) = 0 \ ,
\label{B10}
\end{equation}
\begin{equation}
\frac{\partial}{\partial t}N^{\pm}+\nabla (N^{\pm}{\bf V}^{\pm}) =
0 \ . \label{Cont}
\end{equation}
where \ ${\bf A}$ \ and \ $\varphi $ \ are the EM fields vector
and scalar potentials; \ ${\bf p^{\pm}=\gamma^{\pm}{\bf V}^{\pm}}$
\ is the hydrodynamical momentum of particles, \ ${\bf V}^{\pm}$ \
is velocity, and \ $\gamma ^{\pm}$ \ is a relativistic factor;
$N^{\pm}$ is density in laboratory frame; the ''effective mass''
$G^{\pm}$ depends only on the plasma rest frame density
$n^{\pm}=N^{\pm}/\gamma^{\pm}$ by the following simple relation \
$G^{\pm}=[ 1+( n^{\pm}/n_{c})^{2/3}]^{1/2}$ which is valid for the
arbitrary strength of relativity defined by the ratio
$n^{\pm}/n_c$ .

\bigskip

In what follows we apply the above equations to the problem of the
nonlinear self-guiding of EM beam in highly transparent e-p
plasma. It is convenient to introduce the generalized momentum
$\Pi^{\pm}=G^{\pm}{\bf p}^{\pm}$ and relativistic factor
$\Gamma^{\pm}=G^{\pm}\gamma^{\pm}$. In terms of following
normalized quantities:
\[
\widetilde{t}=\omega t \ ,\qquad
\widetilde{\mathbf{r}}=\frac{\omega}{c}\mathbf{r} \ , \qquad
\widetilde{\mathbf{A}}= \frac{e\mathbf{A}}{m_{e}c^{2}} \ ,\qquad
\widetilde{\varphi }=\frac{e\varphi}{m_{e}c^{2}} \ ,
\]
\[
\widetilde{\mathbf{\Pi}}^{\pm}=\frac{\mathbf{\Pi}^{\pm}}{m_{e}c}\
, \qquad {\widetilde{n}}^{\pm}=\frac{n^{\pm}}{n_0} \qquad \rm{and}
\qquad {{\widetilde{N}}^{\pm}}=\frac{N^{\pm}}{n_0} \ .
\]
Suppressing the tilde, we arrive to the following dimensionless
equations:
\begin{equation}
\frac{\partial^{2}\mathbf{A}}{\partial t^{2}}-\Delta
\mathbf{A+}\frac{\partial }{\partial t}\left(
\mathbf{\nabla}\varphi \right) - \varepsilon^{2}\left(
\mathbf{J}^+ -\mathbf{J}^-\right) =0  \label{B64}
\end{equation}
\begin{equation}
\Delta \varphi =\varepsilon^{2}\left( N^- - N^+\right) \label{B65}
\end{equation}
\begin{equation}
\frac{\partial }{\partial t}\left( \mathbf{\Pi}^{\pm}\mathbf{\pm
A}\right) + \mathbf{\nabla }\left( \Gamma^{\pm}\pm \varphi \right)
= 0 \label{B67}
\end{equation}
\begin{equation}
\frac{\partial N^{\pm}}{\partial t}+\nabla \cdot {\bf J}^{\pm} =0
\label{B69}
\end{equation}
with \ ${\bf J}^{\pm}=n^{\pm}{\bf \Pi}^{\pm}/\Gamma^{\pm}$ \ and \
$\Gamma^{\pm}=\sqrt{(G^{\pm})^{2}+(\mathbf{\Pi }^{\pm})^{2}}$.
Here $\epsilon = \omega_e/\omega \ll 1$, where $\omega$ is the
frequency of EM field and $\omega_e=\sqrt{4\pi e^2n_0/m_e}$. \ The
above equations are similar to those derived for classical
relativistic e-p plasma in \cite{bib:Levan,bib:Ohashi} where the
''effective mass'' was defined by the plasma temperature and
relativistic equation of state, while in the present study \
$G^{\pm}=\sqrt{1+R_{0}^{2}(n^{\pm})^{2/3}}$ \ [with \
$R_{0}=\left( \frac{n_0}{n_{c}}\right)^{1/3}$] which is valid for
entire range of physically allowed densities. Hence, we apply
below the method of multiple scale expansion of the equations in
the small parameter $\epsilon$ \cite{bib:Sun}. We say that all
physical variables $( Q=A,\varphi
,\Pi^{\pm},\Gamma^{\pm},N^{\pm},G^{\pm})$ can be expanded as
\begin{equation}
Q=Q_{\left\{0\right\}}\left( \xi,x_{1,}y_{1,}z_{2}\right)
+\varepsilon Q_{\left\{1\right\}}\left(
\xi,x_{1,}y_{1,}z_{2}\right) \ , \label{B72}
\end{equation}
where \ \ $\left( x_{1},y_{1},z_{2}\right) =$ \ $\left(
\varepsilon x,\varepsilon y,\varepsilon^{2}z\right) $ and
$\xi=z-b\,t$ and $\left( b^{2}-1\right) \sim\varepsilon^{2}$. We
further assume that EM field is circularly polarized
\begin{equation}
\mathbf{A}_{\left\{ 0\perp \right\}
}=\frac{1}{2}(\widehat{\mathbf{x}}+i\widehat{\mathbf{y}}
)A\exp\left( i\xi/b\right)  \label{B73}
\end{equation}
with $A$ being a slowly varying envelope of EM beam.

To the lowest order in \ $\epsilon $ \ , following the standard
procedure,  we get ${\bf \Pi}^{\pm}_{\left\{0\perp\right\}}=\pm
{\bf A}_{\left\{0\perp\right\}}$ \ from transverse component of
Eq.(\ref{B67}), while the first order longitudinal component and
next order Eq.(\ref{B67}) lead to the following: \
$\varphi_{(0)}=0$ \ and  \
$\Gamma^{\pm}_{\left\{0\right\}}=\Gamma_{0} = const$ \ ; thus the
Poisson equation \ $\varphi_{\left\{0\right\}}=0$ \ gives
consequently \ $N^{\pm}_{\left\{0\right\}}=N_{\left\{0\right\}}$ .
Here  \
$\Gamma^{\pm}_{\left\{0\right\}}=\sqrt{G_{\left\{0\right\}}^2+|A|^2}$
\ with \
$G_{\left\{0\right\}}=\sqrt{1+R_{0}^{2}n_{\left\{0\right\}}^{2/3}}$
; since fields vanish at infinity the factor \ $\Gamma
_{0}=\sqrt{1+R_{0}^{2}}\equiv const$. Taking into account that
$n_{\left\{0\right\}}=N_{\left\{0\right\}}/\gamma =
\frac{G_{\left\{0\right\}}}{\Gamma_0}\,N_{\left\{0\right\}}$ \ and
solving algebraic relations \
$\Gamma^{\pm}_{\left\{0\right\}}=\Gamma_{0}$ \ we obtain the
following expression for the plasma density:
\begin{equation}
N_{\left\{0\right\}}=\frac{1}{\delta^3\,\sqrt{1-|{A}|^2/\Gamma_0^2}}
\left( \delta^{2}-|{A}|^2/\Gamma_0^2\right)^{3/2} \ , \label{B122}
\end{equation}
where \ $\delta=R_{0}/\sqrt{ 1+R_{0}^{2}} $ . Note, that
Eq.(\ref{B122}) is valid for arbitrary level of degeneracy
parameter \ $\delta$ \ provided that \ $|A|^2<R_0^2$ .

\bigskip

For slowly varying envelope the Maxwell equation (\ref{B64})
reduces to \ $\nabla_{\perp}^2=\partial_{x_1}^2+\partial_{y_1}^2$:
\begin{equation}
2i\frac{\partial A}{\partial z_{2}}+\nabla _{\perp }^{2}A+\sigma
A-A\frac{ 2N_{\left\{0\right\}}}{\Gamma _{0}}=0  \ , \label{B104}
\end{equation}
here \ $\sigma =(b^{2}-1)/{b^{2}\varepsilon^{2}}$; if we assume
that $b=\omega /{kc}$, than \ $\sigma -1/\Gamma _{0}=0$ is nothing
but the dispersion relation $\omega ^{2}=k^{2}c^{2}+2\Omega
_{e}^{2}$, where $\Omega_e = \omega_e/\sqrt{\Gamma _{0}}$ is a
modified plasma frequency due to degeneracy. Introducing \ $a =
A/R_{0}$ \ , dropping subscripts for the variables $(z_2,x_1,y_1)$
and making self-evident re-normalization: \ $z=2z/\Gamma_{0}$ and
$r_{\perp }=\sqrt{2/\Gamma_{0}}\,r_{\perp }$, the final equation
will read as:
\begin{equation}
2i\frac{\partial a}{\partial z}+\nabla_{\perp }^{2}a+f(\left\vert
a\right\vert ^{2})a=0  \
\label{NLS}
\end{equation}
with
\begin{equation}
f(\left\vert a\right\vert^{2})=1-\frac{\left( 1-\left\vert
a\right\vert^{2}\right)^{3/2}}{\left( 1-\delta^{2}\left\vert
a\right\vert^{2}\right)^{1/2}} \ . \label{NLSF}
\end{equation}
Equation (\ref{NLS}) is a Nonlinear Schr\"odinger Equation (NSE)
with peculiar type nonlinearity function $f$ . The latter is a
growing function of field amplitude $|a|$ attaining its maximum
value $f=1$ at $|a|=1$. For small amplitude case ($|a|\ll 1$) the
nonlinearity function reduces to $f \simeq \beta \,|a|^2$ , where
coefficient $\beta = 0.5(3-\delta^2)$ varies within \ $(1.5-1)$ \
for an arbitrary level plasma degeneracy ($0 < \delta < 1$ ).
While for weak degeneracy level ($\delta \ll 1$) \ $f\simeq 1 -
(1-\left\vert a\right\vert^{2})^{3/2}$ \ and for relativistic
degeneracy ($\delta \to 1$) \ $f\simeq |a|^2$ \ at \ $|a|\leq 1$ .
NSE with various type saturating nonlinearity has been studied
thoroughly in the past
\cite{bib:Tajima,bib:SBM,bib:Vakhitov,bib:MSB_PI,bib:BMS_JPP}; our
eq.(\ref{NLS}) [with (\ref{NLSF})] proves to be similar
qualitatively though showing quantitative difference. First, one
has to establish the existence and the stability of the
self-trapped solutions of Equation (\ref{NLS}). Using the axially
symmetric solution ansatz in the form \ $a=U(r)\,exp(i\lambda z)$
, \ where $r=(x^2+y^2)^{1/2}$ and $\lambda $ is the so called
propagation constant, the ordinary nonlinear differential equation
for the radially dependent envelope $U(r)$ reduces to:
\begin{equation}
\nabla^2_{rr}\,U-\lambda \,U+ f(U^2)\,U = 0 \  \label{NLSU}
\end{equation}
with \ $\nabla^2_{rr} = \frac{d^2}{dr^2}+\frac{1}{r}\frac{d}{dr}$
. We consider the lowest order nodeless (ground state) solution of
above equation with maximum $U_m$ at the center \ ($r=0$) \ and
monotonically decreasing \ ($U\to 0$) \ as \ $r\to \infty$ .
Maximal amplitude $U_m$ is determined by eigenvalue \ $\lambda $ \
which, according to \cite{bib:Vakhitov}, satisfies \ $0<\lambda <
f_m$ , where \ $f_m$ \ is a maximal value of nonlinearity
function.

\begin{figure}
\begin{center}
\includegraphics[scale=0.3,angle=0]{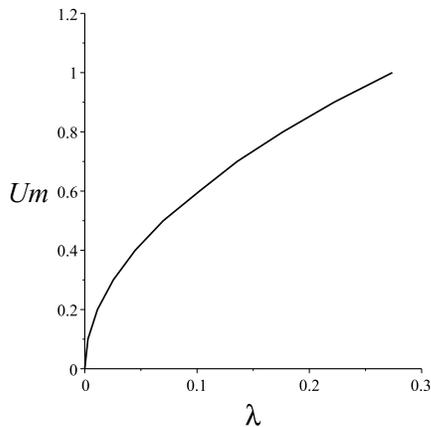}
\caption{ \ \ The typical behavior of soliton amplitude $U_m$
versus propagation constant $\lambda $ for moderate degeneracy
parameter case of $\delta  = 0.5$.} \label{Fig.1.}
\end{center}
\end{figure}

We have performed the numerical simulation study of the
Eq.(\ref{NLSU}) with our nonlinearity (\ref{NLSF}) for arbitrary
level of degeneracy parameter $\delta $. Despite the value of \
$f_m = 1$ \ the allowed range of eigenvalue \ $\lambda $ \ is
significantly narrow \ $\lambda < \lambda_c < 1$ . $U_m$ is a
growing function of \ $\lambda $ \ attaining its maximal value \
$U_m=1$ \ at \ $\lambda = \lambda_c $ . At the same time critical
value $\lambda_c$ decreases with $\delta $ from its maximum being
$0.2912$ at $\delta \ll 1$ to $0.2055$ for ultrarelativistic
degenerate case ($\delta \gg 1$). The typical behavior of soliton
amplitude $U_m$ versus propagation constant $\lambda $ is
exhibited in Fig.1 for moderate degeneracy case $\delta  = 0.5$
while profiles of solitary structure for the variety of $U_m$ is
plotted in Fig.2. Important characteristics of obtained solitary
solutions is a so called beam "power" defined by $P =
2\pi\,\int_0^{\infty}{dr\,r\,U^2(r,\lambda)}$ . Numerical
simulations show that for arbitrary level of degeneracy $\delta $
power \ $P$ \ is a growing function of \ $\lambda \ \rm{and, \
obviously,} \ U_m $ [such behavior for $\delta = 0.5$ is presented
in Fig.3 where \ $P$ \ versus \ $U_m$ \ is plotted]; consequently,
solitary solutions are stable against small perturbations
\cite{bib:Vakhitov}. For \ $\lambda \to 0$ [$U_m \to 0$] \ power
$P\to P_c$ while maximal allowed value of power is achieved at \
$\lambda = \lambda_c \ [U_m=1] $ ; here \ $P_c$ \ is a critical
power. Thus, self-trapped propagation solitary beam could be
formed in such plasmas provided \ $P\geq P_c$ .

\begin{figure}
\begin{center}
\includegraphics[scale=0.32,angle=0]{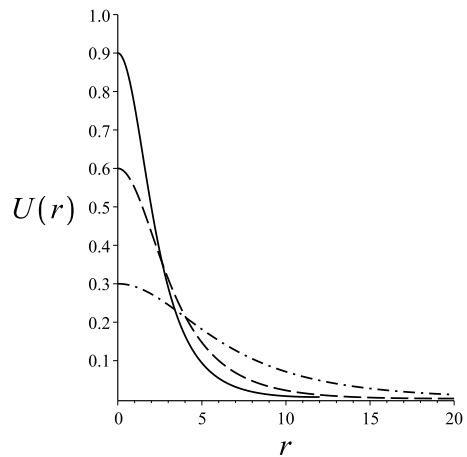}
\caption{ \ \ Profiles of solitary structure for the variety of
$U_m$ .} \label{Fig.2.}
\end{center}
\end{figure}
\begin{figure}
\begin{center}
\includegraphics[scale=0.3,angle=0]{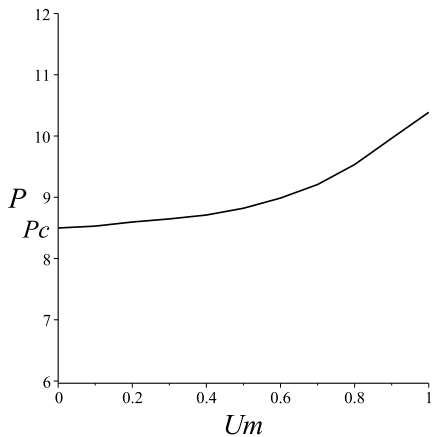}
\caption{ \ \ Power $P$ versus amplitude $U_m$ for $\delta = 0.5$
with critical power $P_c=8.5$.} \label{Fig.3.}
\end{center}
\end{figure}

Critical power of self-trapped beam $P_c$ depends on degeneracy
parameter $\delta $. This relation can be obtained analytically
assuming that for small amplitude case ($\lambda \ll 1, \ f\simeq
\beta U^2 $) with obvious change of variables \ $\left[
U(r)=\sqrt{\frac{\lambda}{\beta}}\,g(\sqrt{\lambda }\,r)\right]$ \
the equation (\ref{NLSU}) reduces to \ $\nabla^2_{rr}\,g-g + g^3 =
0 $ \  \cite{bib:Borisov}, where \ $g(r)$ \ is a stationery Townes
mode. Knowing the Power of ground state Townes mode \
$P_g=2\pi\,\int_0^{\infty}dr\,r\,g^2(r)=11.69$ \ one can find the
critical power to be \ $P_c=\beta^{-1}\,P_g = 23.37/(3-\delta^2)$
. For the nonrelativistically degenerate case $P_c\sim 7.79$ while
for the super-relativistic degenerate case \ $P_c\sim 11.69$.

Critical power in dimensional units can be written in a following
convenient form:
\begin{equation}
[P_c] = 33\ \chi \ \frac{\omega^2}{\Omega_{e}^2}\ \rm{GW} \ ,
\label{PcritD}
\end{equation}
where $\chi = R_0^2/(3+2R_0^2)$. As we mentioned in the
introduction a physically justified range of allowed plasma
densities is presumably within $(10^{24} \div 10^{34})\,cm^{-3}$
\, $[R_0 \sim 1.19\cdot 10^{-2} \div 25.69]$; corresponding
critical power $P_c$ for self-trapped solution to occur is within
$(1.6\cdot 10^{-3} \div
16.5)\,\frac{\omega^2}{\Omega_{e}^2}\,\rm{GW}$ .

\bigskip

At this end it is important to note that recent progress in
creating dense e--p plasmas in Laboratory conditions
\cite{bib:Chen} and achievements in the development of free
electron powerful X-ray sources \cite{bib:Tajima2} indicate that
in the future the generation of the optically thin e–-p plasmas
can be expected with the solid state densities in the range of
$(10^{23}\div 10^{28})\,cm^{-3}$ and above
\cite{bib:Dunne,bib:Shukla-Eliasson}. For X-ray pulse with
wave-length $\sim 3\,nm$ interacting with $n\sim 10^{24}\,cm^{-3}$
density plasma the critical power becomes $\sim 194\,MW$. We would
like to emphasize that in the case of cold non-degenerate
classical plasma (temperature is zero) such effect of the
existence of self-trapped solitary structures is absent.

We performed the direct simulation of derived equations to study
the stability of obtained solitary solutions; analysis of this
study confirms that like in the case of other type saturating
nonlinearities the ground state solution is indeed stable;
moreover, even for Gaussian profile initial radial distribution of
field with power and amplitude close to ground state solution
field quickly relaxes to the equilibrium shape structure;
exception is for the initial profile with amplitude $\sim 1$ or
for the initial amplitude far from ground state solution. In
certain cases when initial Gaussian profile amplitude field is far
from the equilibrium one the amplitude $|A|$ of the evolving field
has a tendency to reach values $\geq 1$ implying that cavitation
(complete expulsion of plasma from field localization area) will
take place. Note, that the possibility of the existence of
cavitation in classical e-p plasma has been demonstrated first
time in \cite{bib:cavitation}. Study of the dynamics of
self-trapped solutions in cavitating regime as well as of the
influence of temporal reshaping of the pulse related with group
velocity dispersion is beyond the scope of present paper.

\bigskip

In conclusion, in present manuscript we showed that in degenerate
e--p plasma the stable self-trapped solitary 2D structures exist
for arbitrary level of degeneracy. We have found the critical
power for the self-guided propagation. The results of given study
can be applied to understand the radiation properties of
astrophysical Gamma-Ray sources as well as may be useful to design
the future laboratory experiments.


{\centerline{\bf * \ * \ *}}

NLS would like to acknowledge the Abdus Salam International Centre
for Theoretical Physics, Trieste, Italy.




\vspace{1cm}

\begin{thebibliography}{}


\bibitem{bib:Tajima} K.A. Holcomb and T. Tajima.
{\it Phys. Rev. D} {\bf 40}, 3809 (1989); V.I. Berezhiani and S.M.
Mahajan.
{\it Phys. Rev. Lett.} {\bf 73}, 1110 (1994);
{\it Phys. Rev. E} {\bf 52}, 1968 (1995).

\bibitem{bib:Stenflo} N.L. Tsintsadze, P.K. Shukla and L. Stenflo.
{\it Eur. Phys. J. D.} {\bf 23}, 109 (2003).

\bibitem{bib:ruffini} W.B. Han, R. Ruffini and S.S. Xue.
{\it Phys. Rev. D} {\bf 86}, 084004 (2012).

\bibitem{bib:aksenov} A.G. Aksenov, R. Ruffini and G.V. Vereshchagin.
{\it Phys. Rev. E} {\bf 81}, 046401 (2010).

\bibitem{bib:Landau} L.D. Landau and E.M. Lifshitz.
{\it Statistical Physics}, Pergamon Press Ltd (1980).

\bibitem{bib:BST_deg} V.I. Berezhiani, N.L. Shatashvili and
N.L. Tsintsadze. {\it Physica Scripta}, {\bf 90(6)}, 068005
(2015).

\bibitem{bib:Katz} J.L. Katz.
{\it ApJ Supplement Series} \ {\bf 127}, 371 (2000).

\bibitem{bib:kennel} A.C.L. Chian and C.F. Kennel. {\it Astrophys. Space Sci.} {\bf 97}, 9
(1983).

\bibitem{bib:Haas} M. McKerr, F. Haas and I. Kourakis. {\it Phys. Rev. E} {\bf 90},
033112 (2014).

\bibitem{bib:Levan} L.N. Tsintsadze, {\it Phys. Plasmas}, {\bf 2}, 4462
(1995).

\bibitem{bib:Ohashi} V.I. Berezhiani,  S.M. Mahajan, Z. Yoshida and
M. Ohhashi.  {\it Phys. Rev. E} {\bf 65}, 047402 (2002).

\bibitem{bib:Sun} G.Z. Sun, E. Ott, Y.C. Lee and P. Guzdar. {\it Phys. Fluids} {\bf 30}, 526
(1987).

\bibitem{bib:Vakhitov} N.G. Vakhitov and A.A. Kolokolov.
{\it Radiophys. Quantum Electron.} {\bf 16}, 783 (1973).

\bibitem{bib:SBM} V. Skarka, V.I. Berezhiani and R. Miklaszewski. {\it  Phys. Rev. E} {\bf
56} 1080 (1997);

\bibitem{bib:MSB_PI} V.I. Berezhiani, S.M.
Mahajan and N.L. Shatashvili. {\it Phys. Rev. A} {\bf 81(5)},
053812 (2010).

\bibitem{bib:BMS_JPP} V.I. Berezhiani, S.M.Mahajan and N.L. Shatashvili.
{\it J. Plasma Phys.} {\bf 76}, 467 (2010).

\bibitem{bib:Borisov} A.B. Borisov, A.V. Borovski, V.V. Korobkin,
A.M. Prokhorov, C.K. Rhodes and O.B. Shiryaev. {\it Sov. Phys.
JETP} {\bf 74(4)}, 604(1992) [{\it Zh.Eksp.Teor.Fiz.} {\bf 101},
1132 (1992)].

\bibitem{bib:Chen} H. Chen, D.D. Meyerhofer, S.C. Wilks, R. Cauble1, F. Dollar, K. Falk, W.
Goldstein, G. Gregori, A. Hazi, E. I. Moses et al., {\it High
Energy Density Physics} {\bf 7}, 225 (2011); G. Sarri, M.E.
Dieckmann, I. Kourakis, A. Di Piazza, B. Reville, C.H. Keitel and
M. Zepf. {\it J. Plasma Physics}, {\bf 81}, 455810401 (2015).

\bibitem{bib:Tajima2} T. Tajima. Plasma Physics Reports, {\bf 29}, 207 (2003),
[Fizika Plazmy, {\bf 29}, 231 (2003)].

\bibitem{bib:Dunne} M. Dunne.
{\it Nature Phys.} {\bf 2}, 2 (2006); \ G.A. Mourou, T. Tajima and
S.V. Bulanov.
{\bf 78}, 309 (2006).

\bibitem{bib:Shukla-Eliasson} Y. Wang, P.K. Shukla and B. Eliasson.
{\it Phys. Plasmas} {\bf 20}, 013103 (2013).

\bibitem{bib:cavitation} L.N. Tsintsadze, {\it Physica Srcipta}, {\bf 50}, 413,
(1994).








\end{thebibliography}
\end{document}